\documentstyle[preprint,eqsecnum,aps,prb]{revtex}
\tighten
\begin{document}
\draft
\title{\Large\bf THE VARIATIONAL THEORY OF THE PERFECT HYPERMOMENTUM FLUID}
\author{O. V. Babourova\cite{bab} and B. N. Frolov\cite{fr}}
\address{Department of Mathematics, Moscow State Pedagogical University,\\
     Krasnoprudnaya 14, Moscow 107140, Russia}
\maketitle
\begin{abstract}
     The variational theory of the perfect hypermomentum fluid is
developed. The new type of the generalized Frenkel condition is considered.
The Lagran\-gi\-an density of such fluid is stated, and the equations of
motion of the fluid and the Weyssenhoff-type evolution equation of the
hypermomentum tensor are derived. The expressions of the matter currents of
the fluid (the canonical energy-momentum 3-form, the metric stress-energy
4-form and the hypermomentum 3-form) are obtained. The special case of the
dilaton-spin fluid with intrinsic spin and dilatonic charge is considered.
\end{abstract}

\section{Introduction}
\label{sec:int}

     The perfect hypermomentum fluid, a fluid element of which is endowed
with an intrinsic hypermomentum, was considered in Refs.\onlinecite{Bab1} and
\onlinecite{Bab:thes}. The variational theory of such fluid in a
metric-affine space $(L_{4},g)$ (Ref.\onlinecite{He:pr}) was developed
by various authors.\cite{Bab:Arg}${}^{-}$\cite{Nov} This theory  generalizes the
variational theory of the Weyssenhoff-Raabe perfect spin fluid based on
accounting the constraints in the Lagrangian density of the fluid with the
help of Lagrange multipliers, which has been developed in case of a
Riemann-Cartan space\cite{Tun1}${}^{-}$\cite{Ol} and in case of a
metric-affine space.\cite{Bab:thes,Bab-Fr} On the other variational methods
of the perfect spin fluid in a Riemann-Cartan space see Refs.\onlinecite
{Min-Kar} and \onlinecite{Kop}.
\par
     The theory of the perfect fluid with intrinsic degrees of freedom
being developed, the additional intrinsic degrees of freedom of a fluid
element are described by the four vectors $\bar{l}_{p}$ ($p = 1,2,3,4$),
called {\it directors}, attached with the each element of the fluid.
Three of the directors ($p = 1,2,3$) are space-like and the fourth one
($p = 4$) is time-like.
\par
     In Riemann and Riemann-Cartan spaces a fluid element endowed with
directors moves according to the Fermi transport that preserves the
orthonormalization of the directors. In a metric-affine space, in which a
metric and a connection are not compatible  it is naturally consider the
directors to be {\it elastic} \cite{Ob-Tr,Bab:tr3,Sm} in the sense that
they can undergo arbitrary deformations during the motion of the fluid.
Nevertheless, in most theories it is accepted that the time-like director is
collinear to the 4-velocity of the fluid element and the orthogonality of
space-like directors to the 4-velocity is maintained.
\par
     The distinction of the variational machinery consists in using the
generalized Frenkel condition,\cite{Bab1,Bab:thes,Bab:Arg}${}^{-}$\cite
{Ob-Tr}
\begin{equation}      \label{eq:01}
J^{\alpha}\!_{\beta} u^{\beta} = 0\; , \qquad J^{\alpha}\!_{\beta}
u_{\alpha} = 0 \; ,
\end{equation}
or the Frenkel condition in its standard classical
form,\cite{Bab:tr3,GR14,BF:Los,Nov}
\begin{equation}   \label{eq:02}
S^{\alpha}\!_{\beta} u^{\beta} = 0 \; ,
\end{equation}
where $J^{\alpha}\!_{\beta}$ and $S^{\alpha}\!_{\beta} := J^{[\alpha}\!_
{\beta ]}$ are the specific intrinsic hypermomentum tensor and the specific
spin tensor of a fluid element, respectively. Another possibility is so
called ``unconstrained hyperfluid''\cite{Ob2}, in which any type of Frenkel
condition is absent.
\par
     In Ref.\onlinecite{Ob2} it is mentioned that in case of generelized
Frenkel condition (\ref{eq:01}) the dilatonic charge  of a fluid element is
expressed in terms of the shear tensor, $J = (4/c^2)\hat J^{\alpha}\!_{\beta}
u_{\alpha}u^{\beta}$ (see decomposition (\ref{eq:11}) in the Sec. \ref{sec:2}).
In this case the hypermomentum fluid can not be of the pure dilationic
type with $\hat J^{\alpha}\!_{\beta} = 0$ and $J \not = 0$. On the other
hand, the Frenkel condition (\ref{eq:02}) leads to the  unusual form of the
evolution equation of the hypermomentum tensor,\cite{GR14,BF:Los} which does
not demonstrate the Weyssenhoff-type dynamics. As to the unconstrained
hyperfluid, in Ref.\onlinecite{Ob2} it is stated that such type of fluid does
not contain the Weyssenhoff spin fluid as a particular case. Therefore all
three kinds of the approaches mentioned are not satisfactory from physical
point of view.
\par
     In this paper we consider the new type of the generalized Frenkel
condition, which allows to construct the hypermomentum perfect fluid theory
with the dilaton-spin fluid and the Weyssenhoff spin fluid as particular
cases. In our approach it is also essential that all four directors
are elastic. None of the orthogonality conditions of the four directors is
maintained during the motion of the fluid. Besides, the time-like director
needs not to be collinear to the 4-velocity of the fluid element. We use the
exterior form variational method according to Trautman\cite{Tr1,Tr2} (see
also Ref.\onlinecite{He:pr}).

\section{The dynamical variables and constraints}
\label{sec:2}
\setcounter{equation}{0}
     In the exterior form language the material frame of the directors
turns into the coframe of 1-forms $l^{p}$ $(p = 1,2,3,4)$, which have dual
3-forms $l_{q}$, while the constraint
\begin{equation}
l^{p} \wedge l_{q} = \delta^{p}_{q}\eta\;, \qquad
l^{p}_{\alpha}l_{p}^{\beta} = \delta_{\alpha}^{\beta}\;, \label{eq:1}
\end{equation}
being fulfilled, where $\eta$ is the volume 4-form and the component
representations are introduced,
\begin{equation}
l^{p} = l^{p}_{\alpha}\theta^{\alpha}\; , \;\;\;\;\;
l_{q} = l_{q}^{\beta}\eta_{\beta}\; .\label{eq:2}
\end{equation}
Here $\theta^{\alpha}$ is a 1-form basis and $\eta_{\beta}$ is a 3-form
defined as \cite{Tr1}
\begin{equation}
\eta_{\beta} = \bar{e}_{\beta}\rfloor \eta = *\theta_{\beta} \; ,
\;\;\;\;\;\;\;\;\;\; \theta^{\alpha} \wedge \eta_{\beta} = \delta^{\alpha}
_{\beta} \eta \; , \label{eq:3}
\end{equation}
where $\rfloor$ means the interior product, $*$ is the Hodge dual operator
and $\bar{e}_{\beta}$ is a basis vector, a coordinate system being
nonholonomic in general.
\par
     Each fluid element possesses a 4-velocity vector $\bar{u}=u^{\alpha}
\bar{e}_{\alpha}$ which is corresponded to a flow 3-form $u$ (Ref.
\onlinecite{Tr2}), $u:=\bar{u}\rfloor\eta=u^{\alpha}\eta_{\alpha}$ and a
velocity 1-form $*\! u=u_{\alpha}\theta^{\alpha}=g(\bar{u},\underline{})$
with
\begin{equation}
*\!u \wedge u = -c^{2}\eta \;, \label{eq:4}
\end{equation}
that means the usual condition $g(\bar{u},\bar{u}) = - c^{2}$, where
$g(\underline{}, \underline{})$ is the metric tensor.
\par
     A fluid element moving, the fluid particles number and entropy
conservation laws are fulfilled,
\begin{eqnarray}
d(n u) = 0 \; , \qquad d(n s u) = 0 \; , \label{eq:7}
\end{eqnarray}
where $n$ is the fluid particles concentration equal to the number of fluid
particles per a volume unit, and $s$ is the the specific (per particle)
entropy of the fluid in the rest frame of reference, respectively.
\par
     The measure of ability of a fluid element to perform the intrinsic
motion is the quantity $\Omega^{q}\!_{p}$ which generalizes the fluid element
``angular velocity'' of the Weyssenhoff spin fluid theory. It has the form
\begin{equation}
\Omega^{q}\!_{p} \eta := u \wedge l^{q}_{\alpha}
{\cal D} l^{\alpha}_{p} \; , \label{eq:9}
\end{equation}
where ${\cal D}$ is the exterior covariant differential with respect to a
connection 1-form $\Gamma^{\alpha}\!_{\beta}$,
\begin{equation}
{\cal D} l^{\alpha}_{p} = d l^{\alpha}_{p} + \Gamma^{\alpha}\!_{\beta}
l^{\beta}_{p} \; . \label{eq:10}
\end{equation}
\par
     An element of the fluid with intrinsic hypermomentum possesses the
additional ``kinetic'' energy 4-form,
\begin{equation}
E =  \frac{1}{2} n J^{p}\!_{q}\Omega^{q}\!_{p}\eta = \frac{1}{2}n
J^{p}\!_{q} u\wedge l^{q}_{\alpha}{\cal D} l^{\alpha}_{p} \; , \label{eq:8}
\end{equation}
where $J^{p}\!_{q}:=J^{\alpha}\!_{\beta} l^{p}_{\alpha} l^{\beta}_{q}$ is the
specific intrinsic hypermomentum tensor representing the new dynamical
quantity which generalizes the spin density of the Weyssenhoff fluid.
\par
     The hypermomentum tensor $J^{p}\!_{q}$ can be decomposed into
irreducible parts,
\begin{eqnarray}
&&J^{p}\!_{q} = \hat J^{p}\!_{q} + \frac{1}{4} \delta^{p}_{q}J\;, \qquad
J := J^{p}\!_{p}\; ,\qquad \hat J^{p}\!_{p} = 0\; ,  \label{eq:11} \\
&& \hat J^{p}\!_{q} := S^{p}\!_{q} + \hat J^{(p}\!_{q)}\; , \qquad
S^{p}\!_{q}:= J^{[p}\!_{q]}\;, \qquad \hat J^{(p}\!_{q)} = J^{(p}\!_{q)}
- \frac{1}{4} \delta^{p}_{q}J\; . \label{eq:112}
\end{eqnarray}
Here $S^{p}\!_{q}$ is the specific spin tensor, $J$ is the specific dilatonic
charge and $\hat J^{(p}\!_{q)}$ is the specific intrinsic proper
hypermomentum (shear) tensor of a fluid element, respectively. We shall name
the quantity $\hat J^{p}\!_{q}$ as the specific {\it traceless hypermomentum
tensor}.
\par
     It is well-known that the spin tensor is spacelike in its nature that
is the fact of fundamental physical meaning. This leads to the classical
Frenkel condition, $S^{\alpha}\!_{\beta} u^{\beta} = 0$. We shall suppose
here that the full traceless part of the hypermomentum tensor
$\hat J^{p}\!_{q}$ (not only the spin tensor but also the tensor
$\hat J^{(p}\!_{q)}$) has such property and therefore satisfies the
generalized Frenkel conditions in the form,
\begin{eqnarray}
&\hat J^{p}\!_{q}u_{p} = 0\; , \qquad u_{p}:= u_{\alpha} l^{ \alpha}_{p}\;,
\label{eq:110}\\
&\hat J^{p}\!_{q} u^{q} = 0\; , \qquad u^{q} := u^{\alpha} l_{\alpha}^{q}\;,
\label{eq:111}
\end{eqnarray}
which can be written in the following way,
\begin{eqnarray}
&\hat J^{p}\!_{q} l_{p}\wedge *\!u   = 0 \; , \label{eq:12} \\
&\hat J^{p}\!_{q} l^{q}\wedge u = 0 \; . \label{eq:120}
\end{eqnarray}
The Frenkel conditions (\ref{eq:110}), (\ref{eq:111}) are equivalent to
the equality,
\begin{equation}
\Pi^{p}_{r} \Pi^{t}_{q} \hat J^{r}\!_{t} = \hat J^{p}\!_{q}\;, \quad
\Pi^{p}_{r} := \delta^{p}_{r} + \frac{1}{c^{2}} u^{p} u_{r}\;. \label{eq:121}
\end{equation}
Here $\Pi^{p}_{r}$ is the projection tensor, which separates the subspace
orthogonal to the fluid velocity.
\par
     The internal energy density of the fluid $\varepsilon$ depends on
the extensive (additive) ther\-mo\-dy\-namic parameters $n$, $s$, $J^{p}\!_
{q}$ and obeys to the first thermodynamic principle,
\begin{equation}
d\varepsilon(n, s, J^{p}\!_{q}) = \frac{\varepsilon + p}{n} dn +
n T ds + \frac{\partial \varepsilon}{\partial J^{p}\!_{q}} dJ^{p}\!_{q}
\; , \label{eq:13}
\end{equation}
where $p$ is the hydrodynamic fluid pressure and $T$ is the temperature.
\par
      We shall consider as independent variables the quantities $n$, $s$,
$J^{p}\!_{q}$, $u$, $l^{q}$, $\theta^{\sigma}$, $\Gamma^{\beta}\!_{\alpha}$,
the constraints (\ref{eq:4}), (\ref{eq:7}), (\ref{eq:12}), (\ref{eq:120})
being taken into account in the Lagrangian density by means of the Lagrange
multipliers.
\par
     In what follows we need the variation,
\begin{equation}
\delta \eta = \eta \frac{1}{2} g^{\alpha\beta}\delta g_{\alpha\beta} +
\delta \theta^{\sigma} \wedge \eta_{\sigma} \;. \label{eq:17}
\end{equation}
As a result of the relation $\theta^{\alpha}\wedge u = u^{\alpha}\eta $
one has,
\begin{equation}
\eta \delta u^{\alpha} = - \delta u \wedge \theta^{\alpha} + \delta \theta^
{\alpha} \wedge u - u^{\alpha}\delta \eta\; . \label{eq:16}
\end{equation}
The relation $*\! u = g_{\alpha\beta}u^{\alpha}\theta^{\beta}$ yields the
variation,
\begin{equation}
\delta *\! u = g_{\alpha\beta}\theta^{\alpha}\delta u^{\beta} +
u^{\alpha}\theta^{\beta}\delta g_{\alpha\beta} + u^{\beta}g_{\sigma\beta}
\delta \theta^{\sigma}\; . \label{eq:161}
\end{equation}
As a result of the resolution of the constraints (\ref{eq:1}) and with the
help of the relations (\ref{eq:3}), one can derive the variations,
\begin{eqnarray}
&&\eta \delta l^{p}_{\alpha} = - \delta \theta^{\sigma}\wedge \eta_{\alpha}
l^{p}_{\sigma} + \delta l^{p} \wedge \eta_{\alpha}\;, \label{eq:14} \\
&&\eta \delta l_{p}^{\alpha} = \delta \theta^{\alpha} \wedge l_{p}
- \delta l^q \wedge l_{q}^{\alpha} l_{p}\; . \label{eq:15}
\end{eqnarray}

\section{The Lagrangian density and the equations of motion of the
     fluid}
\label{sec:3}
\setcounter{equation}{0}
     The perfect fluid Lagrangian density 4-form of the perfect
hypermomentum fluid should be chosen as the remainder after subtraction the
internal energy density of the fluid $\varepsilon$ from the ``kinetic'' energy
(\ref{eq:8}) with regard to the constraints (\ref{eq:4}), (\ref{eq:7}),
(\ref{eq:12}), (\ref{eq:120}) which should be introduced into the Lagrangian
density by means of the Lagrange multipliers $\lambda$, $\varphi$, $\tau$,
$\chi^{q}$, $\zeta_{p}$, respectively. As a result of the previous section
the Lagrangian density 4-form has the form
\begin{eqnarray}
{\cal L}_{m} = L_{m} \eta = - \varepsilon (n, s, J^{p}\!_{q}) \eta +
\frac{1}{2}n J^{p}\!_{q} u\wedge l^{q}_{\alpha}{\cal D} l^{\alpha}_{p}
+ n u\wedge d\varphi + n\tau u \wedge ds \nonumber\\
+ n \lambda (*\!u \wedge u + c^2 \eta) + n \chi^q\hat J^p\!_q l_p\wedge *\!u
+ n\zeta_p \hat J^p\!_q l^q\wedge u\;. \label{eq:20}
\end{eqnarray}
\par
     The fluid motion equations and the evolution equation of  the
hypermomentum tensor are derived by the variation of (\ref{eq:20}) with
respect to the independent variables $n$, $s$, $J^{p}\!_{q}$, $u$,
$l^{q}$, and the Lagrange multipliers, the thermodynamic principle
(\ref{eq:13}) being taken into account.  We shall consider the 1-form  $l^q$
as an  independent  variable  and the 3-form $l_p$ as a function of $l^q$ by
means of (\ref{eq:1}). As a result of such variational machinery one gets the
constraints (\ref{eq:4}), (\ref{eq:7}), (\ref{eq:12}), (\ref{eq:120}) and the
following variational equations,
\begin{eqnarray}
\delta n : &&\quad (\varepsilon + p) \eta - \frac{1}{2}n
J^{p}\!_{q} u\wedge l^{q}_{\alpha}{\cal D} l^{\alpha}_{p}
- n u\wedge d\varphi = 0 \; ,\label{eq:21}\\
\delta s : &&\quad  T \eta + u \wedge d\tau = 0 \;, \label{eq:22}\\
\delta J^{p}\!_{q} : &&\quad \frac{\partial \varepsilon}{\partial J^{p}\!_{q}}
= \frac{1}{2}n\Omega^{q}\!_{p} - n(\chi^{q}u_{p} - \zeta_{p}u^{q}) +
\frac{1}{4}n\delta^{p}_{q} (\chi^{r}u_{r} - \zeta_{r}u^{r})\;,\label{eq:23}\\
\delta u : &&\quad d\varphi + \tau ds - 2\lambda *\! u + \chi^{q}\hat
J_{\beta q}\theta^{\beta} - \zeta_{p}\hat J^{p}\!_{q} l^{q}
+ \frac{1}{2}J^{p}\!_{q} l^{q}_{\alpha} {\cal D}l^{\alpha}_{p} = 0\; ,
\label{eq:25} \\
\delta l^{q} : &&\quad \frac{1}{2}\dot{J}^{\sigma}\!_{\rho}l_{q}^{\rho}
\eta_{\sigma} - \chi^{r}\hat J^{p}\!_{r} u_q l_p
 - \zeta_{r} \hat J^{r}\!_{q} u = 0\;. \label{eq:26}
\end{eqnarray}
Here the ``dot'' notation for the tensor object $\Phi$ is introduced,
\begin{equation}
\dot{\Phi}^{\alpha}\!_{\beta} := *\!(u\wedge {\cal D}\Phi^{\alpha}\!_
{\beta})\; . \label{eq:27}
\end{equation}
\par
     Multiplying the equation (\ref{eq:25}) by $u$ from the left externally
and using (\ref{eq:7}) and (\ref{eq:21}), one derives the expression for the
Lagrange multiplier $\lambda$,
\begin{equation}
2 n c^{2} \lambda = \varepsilon + p \; . \label{eq:28}
\end{equation}
\par
     As a consequence of the equation (\ref{eq:21}) and the constraints
(\ref{eq:4}), (\ref{eq:7}), (\ref{eq:12}), (\ref{eq:120}) one can verify that
the Lagrangian density 4-form (\ref{eq:20}) is proportional to the
hydrodynamic fluid pressure, ${\cal L}_{m} = p \eta$, which corresponds to
Ref.\onlinecite{Rit}.

\section{The evolution equation of the hypermomentum tensor}
\label{sec:4}
\setcounter{equation}{0}
     The variational equation (\ref{eq:26}) represents the evolution equation
of the hypermomentum tensor. Multiplying the equation (\ref{eq:26}) by
$l^{p}_{\beta}\theta^{\alpha}\wedge\dots$ from the left externally one gets,
\begin{equation}
\frac{1}{2}\dot{J}^{\alpha}\!_{\beta} - \chi^{r}\hat J^{\alpha}\!_{r}
u_{\beta} - \zeta_{r}\hat J^{r}\!_{\beta}u^{\alpha} = 0\; . \label{eq:30}
\end{equation}
Contractions (\ref{eq:30}) with $u_{\alpha}$ and then with $u^{\beta}$ yield
the expressions for the Lagrange multipliers,
\begin{eqnarray}
&& \zeta_{r}\hat J^{r}\!_{\beta} = - \frac{1}{2c^2} \dot{J}^{\gamma}
\!_{\beta} u_{\gamma}\; , \label{eq:31} \\
&& \chi^{r}\hat J^{\alpha}\!_{r} = - \frac{1}{2c^2} \dot{J}^{\alpha}\!_
{\gamma}u^{\gamma}\; . \label{eq:32}
\end{eqnarray}
After the substitution of (\ref{eq:31}) and (\ref{eq:32}) into (\ref{eq:30})
one gets the evolution equation of the hypermomentum tensor,
\begin{equation}
\dot{J}^{\alpha}\!_{\beta} + \frac{1}{c^{2}} \dot {J}^{\alpha}\!_{\gamma}
u^{\gamma}u_\beta + \frac{1}{c^{2}} \dot {J}^{\gamma}\!_{\beta}
u_{\gamma}u^{\alpha} = 0\; . \label{eq:33}
\end{equation}
This equation generalizes the evolution equation of the spin tensor in the
Weyssenhoff fluid theory.
\par
     The equation (\ref{eq:33}) has the consequence,
\begin{equation}
\dot {J}^{\alpha}\!_{\beta} u_{\alpha} u^{\beta} = 0\; , \label{eq:34}
\end{equation}
which permits to represent the evolution equation of the hypermomentum tensor
(\ref{eq:33}) in the form,
\begin{equation}
\Pi^{\alpha}_{\sigma}\Pi^{\rho}_{\beta} \dot{J}^{\sigma}\!_{\rho} = 0 \;,
\label{eq:355}
\end{equation}
where the projection tensor $\Pi^{\alpha}_{\sigma}$ has been defined in
(\ref{eq:121}). The evolution equation of the hypermomentum tensor in the
form (\ref{eq:355}) was derived in Ref. \onlinecite{Nov}.
\par
     The contraction (\ref{eq:33}) on the indices $\alpha$ and $\beta$
gives with the help of (\ref{eq:34}) the dilatonic charge conservation law,
\begin{equation}
\dot{J} = 0\; . \label{eq:361}
\end{equation}

\section{The energy-momentum tensor of the perfect hypermomentum
     fluid}
\label{sec:5}
\setcounter{equation}{0}
     With the help of the matter Lagrangian density (\ref{eq:20}) one can
derive the external matter currents which are the sources of the
gravitational field. In case of the perfect hypermomentum fluid the matter
currents are the canonical energy-momentum 3-form $\Sigma_{\sigma}$,
the metric stress-energy 4-form $\sigma^{\alpha\beta}$  and the
hypermomentum 3-form ${\cal J}^{\alpha}\!_{\beta}$, which are determined as
variational derivatives.
\par
     The variational derivative of the explicit form of the Lagrangian
density (\ref{eq:20}) with respect to $\theta^{\sigma}$ yields the canonical
energy-momentum 3-form,
\begin{equation}
\Sigma_{\sigma} := \frac{\delta{\cal L}_{m}}{\delta \theta^{\sigma}}
 = - \varepsilon\eta_{\sigma} + 2\lambda n u_{\sigma} u
+ 2 c^{2}\lambda n \eta_{\sigma} - n \chi^{r}\hat J^{q}\!_{r}
(g_{\sigma\rho}l^{\rho}_{q} u + u_{\sigma}l_{q})
+ \frac{1}{2}n \dot{J}^{\rho}\!_{\sigma}\eta_{\rho}\; . \label{eq:350}
\end{equation}
Using the explicit form of the Lagrange multiplier (\ref{eq:28}), one gets,
\begin{equation}
\Sigma_{\sigma} =  p \eta_{\sigma} + \frac{1}{c^{2}}(\varepsilon + p)
u_{\sigma} u + \frac{1}{2}n \dot{J}^{\rho}\!_{\sigma}\eta_{\rho}
- n \chi^{r}\hat J^{\rho}\!_{r}(g_{\sigma\rho} u +
l^{p}_{\rho}l_{p}u_{\sigma})\; . \label{eq:351}
\end{equation}
On the basis of the evolution equation of the hypermomentum tensor
(\ref{eq:33}) and with the help of (\ref{eq:32}) the expression (\ref{eq:351})
reads,
\begin{equation}
\Sigma_{\sigma} = p\eta_{\sigma} + \frac{1}{c^{2}}(\varepsilon + p) u_
{\sigma} u  + \frac{1}{c^{2}}n g_{\alpha [\sigma}\dot{J}^{\alpha}\!_
{\beta]}u^{\beta}u \; . \label{eq:36}
\end{equation}
After some algebra one can get the other form of the canonical
energy-momentum 3-form,
\begin{equation}
\Sigma_{\sigma} =  p \eta_{\sigma} + \frac{1}{c^{2}}(\varepsilon + p)
u_{\sigma} u + \frac{1}{c^{2}}n \dot{S}_{\sigma}\!_{\beta} u^{\beta} u
- \frac{1}{c^{2}} n J^{\beta}\!_{[\sigma} Q_{\alpha ]\beta \gamma}
u^{\gamma} u^{\alpha} u \; , \label{eq:37}
\end{equation}
where $Q_{\alpha \beta \gamma}$ are components of a nonmetricity 1-form,
\begin{equation}
{\cal Q}_{\alpha \beta } := - {\cal D} g_{\alpha \beta } =
Q_{\alpha \beta \gamma} \theta^{\gamma}\; . \label{eq:nem}
\end{equation}
\par
     The metric stress-energy 4-form can be derived in the same way,
\begin{eqnarray}
&& \sigma^{\alpha\beta} := 2\frac{\delta {\cal L}_{m}}{\delta g_{\alpha\beta}}
= T^{\alpha\beta} \eta \; , \nonumber \\
&& T^{\alpha\beta} = - \varepsilon g^{\alpha\beta} + 2 n \lambda (u^{\alpha}
u^{\beta} + c^{2}g^{\alpha\beta}) - 2 n \chi^{r} \hat J^{(\alpha}\!_{r}
u^{\beta )} \nonumber \\
&& = p g^{\alpha\beta} + \frac{1}{c^{2}}(\varepsilon + p)u^{\alpha}u^{\beta}
+ \frac{1}{c^{2}}n \dot{J}^{(\alpha}\!_{\gamma} u^{\beta )} u^{\gamma} \; .
\label{eq:35}
\end{eqnarray}
\par
     For the hypermomentum 3-form one finds
\begin{equation}
{\cal J}^{\alpha}\!_{\beta} := - \frac{\delta{\cal L}_{m}}{\delta\Gamma^
{\beta}\!_{\alpha}} = \frac{1}{2} n J^{\alpha}\!_{\beta} u \; .
\label{eq:38}
\end{equation}
\par
     The expressions of the canonical energy-momentum 3-form
(\ref{eq:36}), the metric stress-energy 4-form (\ref{eq:35}) and the
hypermomentum 3-form (\ref{eq:38}) are compatible in the sense that they
satisfy to the Noether identity,
\begin{equation}
\theta^{\alpha} \wedge \Sigma_{\beta} = \sigma^{\alpha}\!_{\beta} +
{\cal D} {\cal J}^{\alpha}\!_{\beta} \; , \label{eq:39}
\end{equation}
that corresponds to the $GL(n,R)$-invariance of the Lagrangian density
(\ref{eq:20}).\cite{He:pr}
\par
     Let us consider the special case of the perfect spin fluid with dilatonic
charge, a fluid element of which does not possess the specific shear momentum
tensor, $\hat J^{(p}\!_{q)} = 0$, and is endowed only with the specific spin
momentum tensor $S^{p}\!_{q}$ and the specific dilatonic charge $J$. In this
case the canonical energy-momentum 3-form (\ref{eq:36}) reads,
\begin{equation}
\Sigma_{\sigma} = p\eta_{\sigma} + \frac{1}{c^{2}}(\varepsilon + p) u_
{\sigma} u  + \frac{1}{c^{2}}n g_{\alpha [\sigma}\dot{S}^{\alpha}\!_
{\beta]}u^{\beta}u \; , \label{eq:40}
\end{equation}
where the specific energy density $\varepsilon$ contains the energy density
of the  dilatonic  interaction of the fluid. If the dilatonic charge
also vanishes,  $J  = 0$, then the expression (\ref{eq:40}) will describe
the canonical energy-momentum 3-form of the Weyssenhoff perfect  spin  fluid
in a metric-affine space.

\section{Conclusions}
     The essential feature of the constructed variational theory of the
hypermomentum perfect fluid is the assumption that the frame realized by
all four directors is elastic. The deformation of the directors during the
motion of the fluid element, from one side, generates the space-time
nonmetricity and, from the other side, allows nonmetricity of the
space-time to be discovered. As the consequence of this fact the Lagrangian
density (\ref{eq:20}) does not contain the term maintaining the orthogonality
of the directors. The time-like director needs not to be collinear to the
4-velocity of the fluid element. The essential feature of our variational
approach is the using the Frenkel conditions (\ref{eq:110}), (\ref{eq:111}),
which do not coincide nor with their classical form, when the Frenkel
condition was imposed on the spin tensor, $S^{\alpha}\!_{\beta} u^{\beta}=0$,
nor with its generalized form, when the Frenkel condition was imposed on the
full intrinsic hypermomentum tensor, $J^{\alpha}\!_{\beta} u^{\beta} =
J^{\alpha}\!_{\beta} u_{\alpha} = 0$.
\par
     We have derived the expression for the energy-momentum tensor of the
fluid (\ref{eq:36}), which coincides with one that has been obtained earlier.
\cite{Ob-Tr,Bab:tr3} But our approach does not contain the
shortcomings, which are inherent in the previous theories of the
hypermomentum perfect fluid. First of all, our variational theory contains
the Weyssenhoff spin fluid as the particular case that is important from
physical point of view. Then, the evolution equation of the hypermomentum
tensor (\ref{eq:33}) demonstrates the Weyssenhoff-type dynamics. At last,
the hypermomentum perfect fluid theory developed allows to describe as the
special case the perfect spin fluid with dilatonic charge. It should be
important to investigate the consequences of the employing the perfect fluid
of such type as the gravitational field source in cosmological and
astrophysical problems. For example, it is interesting to clarify whether
the corresponding field equations have the regular solution with the upper
limit for $\varepsilon$ (the limiting energy density of the fluid).

\acknowledgements

     This paper is partly supported by the scientific programm
``Univesitety Rossii''.

\end{document}